\documentclass[fleqn,usenatbib]{mnras}

\usepackage{newtxtext,newtxmath}
\usepackage[T1]{fontenc}
\usepackage{ae,aecompl}
\usepackage{booktabs}

\usepackage{rotfloat}
\usepackage[tableposition=top]{caption}
\usepackage{pdflscape}

\usepackage{graphicx}	% Including figure files
\usepackage{amsmath}	% Advanced maths commands
\usepackage{amssymb}	% Extra maths symbols

\title[Wide binary metallicities]{Wide Binaries in Tycho-{\it Gaia} II: Metallicities, Abundances, and Prospects for Chemical Tagging}

\author[J. J. Andrews et al.]{
Jeff J.~Andrews,$^{1,2}$\thanks{E-mail: andrews@physics.uoc.gr}
Julio Chanam\'e,$^{3,4}$
Marcel A.~Ag\"ueros$^{5}$
\\
$^1$ Foundation for Research and Technology -- Hellas, IESL, Voutes, 71110 Heraklion, Greece \\
$^2$ Physics Department \& Institute of Theoretical \& Computational Physics, University of Crete, 71003 Heraklion, Crete, Greece \\
$^3$ Instituto de Astrof\'isica, Pontificia Universidad Cat\'olica de Chile, Av.~Vicu\~na Mackenna 4860, 782-0436 Macul, Santiago, Chile \\
$^4$ Millennium Institute of Astrophysics, Santiago, Chile \\
$^5$ Department of Astronomy, Columbia University, 550 West 120th Street, New York, NY 10027, USA
}

\date{Accepted XXX. Received YYY; in original form ZZZ}

\pubyear{2017}

\begin{document}
\label{firstpage}
\pagerange{\pageref{firstpage}--\pageref{lastpage}}
\maketitle

\begin{abstract}
From our recent catalog based on the first {\it Gaia} data release (TGAS), we select wide binaries in which both stars have been observed by the Radial Velocity Experiment (RAVE) or the Large Sky Area Multi-Object Fibre Spectroscopic Telescope (LAMOST). Using RAVE and LAMOST metallicities and RAVE Mg, Al, Si, Ti, and Fe abundances, we find that the differences in the metallicities and elemental abundances of components of wide binaries are consistent with being due to observational uncertainties, in agreement with previous results for smaller and more restricted samples. The metallicity and elemental abundance consistency between wide binary components presented in this work confirms their common origin and bolsters the status of wide binaries as ``mini-open clusters.'' Furthermore, this is evidence that wide binaries are effectively co-eval and co-chemical, supporting their use for e.g., constraining age-activity-rotation relations, the initial-final mass relation for white dwarfs, and M-dwarf metallicity indicators. Additionally, we demonstrate that the common proper motion, common parallax pairs in TGAS with the most extreme separations (s $\gtrsim$ 0.1 pc) typically have inconsistent metallicities, radial velocities or both and are therefore likely to be predominantly comprised of random alignments of unassociated stars with similar astrometry, in agreement with our previous results.  Finally, we propose that wide binaries form an ideal data set with which to test chemical tagging as a method to identify stars of common origin, particularly because the stars in wide binaries span a wide range of metallicities, much wider than that spanned by nearby open clusters.
\end{abstract}


%%%%%%%%%%%%%%%%%%%%%%%%%%%%%%%%%%%%%%%%%%%%%%%%%%

%%%%%%%%%%%%%%%%% BODY OF PAPER %%%%%%%%%%%%%%%%%%

\section{Introduction}
\label{sec:intro}

The orbital separation distribution of stellar binaries spans from $\sim$10 R$_{\odot}$ to $\sim$pc \citep{chaname04, quinn09b, raghavan10}. Still, binaries are expected to have formed together from the same pre-stellar clouds in the case of close binaries \citep{bonnell94a,bonnell94b,bate95}, and from pairs of stars that associate during the dissolution of low-mass stellar clusters in the case of wide binaries \citep{kouwenhoven10,moeckel11}. If wide binaries form from the disrupted remnants of low mass stellar clusters, their components should have differences in chemical abundances and ages similar to the dispersion observed today in stellar clusters.

While other authors have proposed more exotic formation scenarios \citep{reipurth12,tokovinin17}, these scenarios all imply that the components of wide binaries have essentially identical elemental abundances and ages, a characteristic which enables wide binaries to provide unique leverage on difficult-to-constrain stellar physics \citep[for a review, see][]{soderblom10} such as the initial-final mass relation for white dwarfs \citep{finley97,catalan08,zhao12,andrews15}, age-activity-rotation relations \citep{soderblom91, barnes07,garces11,chaname12}, and metallicity indicators \citep{bonfils05,lepine07,rojas_ayala10,li14}.

Once spectroscopic measurements began making the comparisons, it was realized that open cluster members tend to have not only the same metallicities, but also, with rare exceptions (e.g., red giants), the same detailed chemical abundances \citep{cayrel85, gratton94, schuler03, de_silva07, bovy16}. As modern spectroscopic surveys such as those conducted by the Radial Velocity Experiment \citep[RAVE;][]{kunder17}, the Sloan Extension for Galactic Understanding and Exploration \citep[SEGUE;][]{yanny09}, the Large Sky Area Multi-Object Fibre Spectroscopic Telescope \citep[LAMOST;][]{luo15}, the Apache Point Observatory Galactic Evolution Experiment  \citep[APOGEE;][]{holtzman15}, {\it Gaia}-ESO \citep{gilmore12}, and GALactic Archaeology with HERMES \citep[GALAH;][]{de_silva15} plan to derive, or are already deriving, abundance patterns of multiple elements for large numbers of stars, the remarkable abundance consistency of open cluster stars is being confirmed. For instance, using APOGEE spectra of 90 stars in seven open clusters, \citet{ness17} recently concluded that members of each cluster closely follow each cluster's unique chemical abundance signature for 20 elements.

Chemical tagging attempts to exploit the unique chemical signatures of stellar structures to identify stellar populations, such as globular clusters and infalling satellite galaxies, that have long since been disrupted by the Milky Way \citep{freeman02}. Previous tests of chemical tagging \citep{bovy16,ness17} have been promising, but focused on open clusters \citep[although see][who use an alternative test, c.f. Section \ref{sec:chemical_tagging}]{hogg16}. However, as first argued by \citet{binney87}, wide binaries are in essence the smallest expressions of open clusters. If wide binaries are in fact formed from either different fragments of the same cloud or in co-chemical environments, they should provide as good a test of the ability of chemical tagging to identify associated stellar structures in the Galaxy. For instance, \citet{kraus09} found that components of wide binaries in the Taurus-Auriga system have ages more consistent than randomly paired stars within the same system. If the same trend is true with metallicities and abundances, then the consistency of wide binary metallicities and elemental abundances provides a limit for the capabilities of chemical tagging.

Analysis of the specific elemental abundances of wide binaries components has lagged behind corresponding work for open clusters. \citet{gizis97} first showed that three M dwarfs have metallicities similar to their wide companions. \citet{gratton01} used a sample of six wide binaries with similar components to demonstrate that the metallicities of wide binaries are consistent. Later studies have confirmed these conclusions with slightly larger samples \citep{desidera04, desidera06b}. Combined, these studies include over 50 pairs and show that the metallicities of wide binary components are typically consistent to within 0.02 dex.

Analysis of the detailed abundance patterns (beyond Fe) of wide binary components has been limited to V \citep{desidera06b} and Li \citep{martin02} and a handful of additional wide binaries, particularly those known to host extrasolar planets. These include the X0-2N/X0-2S system \citep{teske13, teske15}, 16 Cyg A/B \citep{laws01, ramirez11, schuler11, tucci14}, HAT-P-1 \citep{liu14}, and HD 20872/20871 \citep{mack14}. Separately, in a detailed chemical analysis of over 1600 stars, \citet{brewer16} discuss the abundance differences of nine binary systems. In agreement with studies of metallicity, these systems tend to show nearly identical abundances for a range of elements, although small but significant differences do exist. However, note that \citet{oh17b} recently analysed a wide binary that shows abundance differences of refractory elements larger than 0.2 dex.

These studies of metallicity and abundance patterns in wide binaries have been limited to relatively small numbers of specifically selected pairs, generally with comparable-mass components \citep[e.g., the potential solar-type planet hosts of][]{desidera04,desidera06b}. Using the Tycho-{\it Gaia} Astrometric Solution (TGAS) catalog \citep{gaia_DR_1,gaia_DR1_2}, we recently identified $>$7000 wide binaries \citep[][hereafter Paper I]{andrews17}. A subset of these pairs has been observed by either RAVE \citep[]{kunder17} or LAMOST \citep{luo15}, each of which provide radial velocities (RVs) and metallicities. Additionally, RAVE provides elemental abundances for Mg, Al, Si, Ti, and Fe. In this work, we extend previous analyses comparing the metallicities of the components of wide binaries to a larger and more diverse sample, and we compare the abundances of several elements.

In Section \ref{sec:sample}, we describe how our population is both an extension of and complement to previous binary samples. In Section \ref{sec:metallicity}, we compare the metallicities of wide binary components, and we provide the analogous comparison for specific elemental abundances in Section \ref{sec:abundance}. In Section \ref{sec:discussion} we discuss the implications for the level of metallicity and elemental consistency of the stars in wide binaries, and we explore the possibility of using wide binaries for studies of chemical tagging in Section \ref{sec:chemical_tagging}. We conclude in Section \ref{sec:conclusions}.

\section{Our Sample of Wide Binaries}
\label{sec:sample}
From our catalog of 7108 wide binaries identified in TGAS in Paper I, we select the subset of pairs in which the posterior probability is above 99\% when using a power-law prior on the distribution of orbital separations (for details, see Paper I). To further reduce contamination from randomly aligned stellar pairs, we select only those pairs with projected separations less than 4$\times$10$^4$ AU, based on our contamination estimates from Paper I. This restriction winnows our sample down to 4491 binaries. Estimates using a synthetic catalog of randomly aligned stellar pairs indicates that the contamination rate of these pairs should be 5-10\%. We first cross-match these pairs with the RAVE DR5 catalog \citep{kunder17} and obtain 115 wide binaries in which both components are detected by RAVE and in which the fitting algorithm successfully converged ({\tt ALGO\_CONV = 0}). The {\it Gaia} Source IDs, RVs, metallicities, surface gravities and effective temperatures for the stars in each of these 115 pairs are provided in Table \ref{tab:RAVE}.

\restylefloat{table}
\begin{sidewaystable}
\centering
% \vspace{-3in}  % Depending on column and page
\vspace{3in}  % Depending on column and page
\caption{The first 10 wide binaries from the 115 binaries in which RAVE has observed both stars in the pair and data pass the quality cuts described in Section \ref{sec:sample}. We provide Source IDs from {\it Gaia}, RVs and metallicities (along with their corresponding uncertainties), surface gravities, and effective temperatures as measured by RAVE. \label{tab:RAVE}} 
\resizebox{\textwidth}{!}{
\begin{tabular}{cccccccccccccc}
\toprule
Source ID & RV$_1$ & $\sigma_{\rm RV, 1}$ & [M/H]$_1$ & $\sigma_{\rm [M/H], 1}$& T$_{\rm eff, 1}$ & log $g_1$ & Source ID & RV$_2$ & $\sigma_{\rm RV, 2}$ & [M/H]$_2$& $\sigma_{\rm [M/H], 2}$ & T$_{\rm eff, 2}$ & log $g_2$ \\
 & (km s$^{-1}$) & (km s$^{-1}$) &  &  & (K) &  &  & (km s$^{-1}$) & (km s$^{-1}$) &  &  & (K) & \\
\midrule
5061131359288965120 & 29.9 & 0.7 & -0.38 & 0.09 & 6200 & 4.07 & 5061131428008441856 & 28.8 & 1.2 & -0.44 & 0.11 & 5590 & 3.83 \\
6344178312696928256 & 16.4 & 7.6 & 0.83 & 0.29 & 5250 & 4.00 & 6344178312697829376 & 2.8 & 0.9 & -0.02 & 0.09 & 5690 & 4.02 \\
6347160050791913472 & 67.9 & 1.1 & 0.05 & 0.09 & 5910 & 4.04 & 6347160050793576448 & 66.8 & 0.9 & 0.07 & 0.10 & 5510 & 4.12 \\
6350439275502832640 & -5.7 & 1.3 & -0.37 & 0.09 & 4770 & 3.95 & 6350439309862571008 & -3.6 & 0.7 & -0.42 & 0.08 & 5750 & 4.53 \\
6355365431192546304 & -3.0 & 0.7 & -0.27 & 0.09 & 5940 & 3.97 & 6355365602991238144 & 65.8 & 1.3 & -0.18 & 0.12 & 6140 & 4.58 \\
6358484505161891840 & -18.0 & 0.7 & 0.00 & 0.09 & 5940 & 4.11 & 6358484505163066368 & -19.0 & 1.6 & 0.23 & 0.12 & 6640 & 4.52 \\
6361299358008869888 & -13.9 & 0.8 & -0.06 & 0.10 & 5180 & 3.56 & 6361299598527038464 & -9.9 & 1.7 & -0.37 & 0.12 & 6330 & 4.04 \\
4787486657714644992 & -19.0 & 0.8 & -0.22 & 0.09 & 5620 & 3.63 & 4787486726434928640 & -16.2 & 2.5 & -0.25 & 0.10 & 5550 & 3.58 \\
5512742944215410688 & 31.8 & 0.8 & -0.16 & 0.09 & 4800 & 2.18 & 5512743012934887424 & 32.1 & 1.1 & -0.00 & 0.09 & 4700 & 1.67 \\
5212670587315161088 & -5.4 & 2.2 & 0.28 & 0.12 & 7240 & 4.04 & 5212670621675047936 & 2.1 & 0.7 & -0.43 & 0.11 & 5950 & 3.85 \\
\bottomrule
\end{tabular}
}
\end{sidewaystable}

\restylefloat{table}
\begin{sidewaystable}
\centering
% \vspace{-3in}  % Depending on column and page
\vspace{3in}  % Depending on column and page
\caption{The first 10 wide binaries from the 62 binaries in which LAMOST has observed both stars in the pair and data pass the quality cuts described in Section \ref{sec:sample}. The data provided are analogous to that from Table \ref{tab:RAVE}. \label{tab:LAMOST}} 
\resizebox{\textwidth}{!}{
\begin{tabular}{cccccccccccccc}
\toprule
Source ID & RV$_1$ & $\sigma_{\rm RV, 1}$ & [Fe/H]$_1$ & $\sigma_{\rm [Fe/H], 1}$& T$_{\rm eff, 1}$ & log $g_1$ & Source ID & RV$_2$ & $\sigma_{\rm RV, 2}$ & [Fe/H]$_2$& $\sigma_{\rm [Fe/H], 2}$ & T$_{\rm eff, 2}$ & log $g_2$ \\
 & (km s$^{-1}$) & (km s$^{-1}$) &  &  & (K) &  &  & (km s$^{-1}$) & (km s$^{-1}$) &  &  & (K) & \\
\midrule
361617623345021952 & -10.5 & 1.4 & -0.42 & 0.01 & 5510 & 4.61 & 361620990599379072 & -8.4 & 3.5 & -0.55 & 0.03 & 6170 & 4.15 \\
370238447342099968 & 3.8 & 4.1 & 0.08 & 0.04 & 4850 & 4.40 & 370243154626255104 & 1.5 & 0.7 & 0.02 & <0.01 & 6130 & 4.36 \\
681192464665370112 & -2.6 & 2.0 & 0.16 & 0.02 & 5900 & 4.36 & 681192464665370240 & -1.1 & 2.1 & 0.10 & 0.02 & 5940 & 4.38 \\
2129614316308043776 & -8.4 & 0.6 & -0.18 & <0.01 & 6650 & 4.18 & 2129614316308043776 & -8.9 & 1.0 & -0.03 & <0.01 & 5990 & 4.48 \\
3131870017437089792 & -6.0 & 2.6 & 0.34 & 0.03 & 5890 & 4.25 & 3131870223595520000 & 39.3 & 1.3 & -0.24 & 0.01 & 6340 & 4.25 \\
3135687315651918848 & 54.8 & 1.8 & -0.15 & 0.02 & 6570 & 4.25 & 3135687350011657216 & 50.0 & 2.5 & -0.07 & 0.02 & 6340 & 4.19 \\
3020220658951548416 & 39.4 & 1.9 & -0.08 & 0.02 & 6420 & 4.30 & 3020220865109978624 & 42.9 & 11.8 & -0.10 & 0.14 & 6050 & 4.35 \\
3351402700928655360 & 54.1 & 4.7 & -0.05 & 0.04 & 5310 & 4.69 & 3351402700928655360 & 53.1 & 2.5 & -0.21 & 0.03 & 6140 & 4.36 \\
804544956197775616 & -13.0 & 1.0 & 0.08 & <0.01 & 6090 & 4.38 & 804547876775536384 & -17.9 & 0.9 & 0.00 & <0.01 & 7300 & 4.00 \\
817365158699243904 & 24.2 & 1.8 & 0.16 & 0.01 & 5460 & 4.62 & 817365227418720512 & 23.9 & 0.9 & 0.06 & <0.01 & 5810 & 4.45 \\
\bottomrule
\end{tabular}
}
\end{sidewaystable}

\begin{figure}
	\includegraphics[width=\columnwidth]{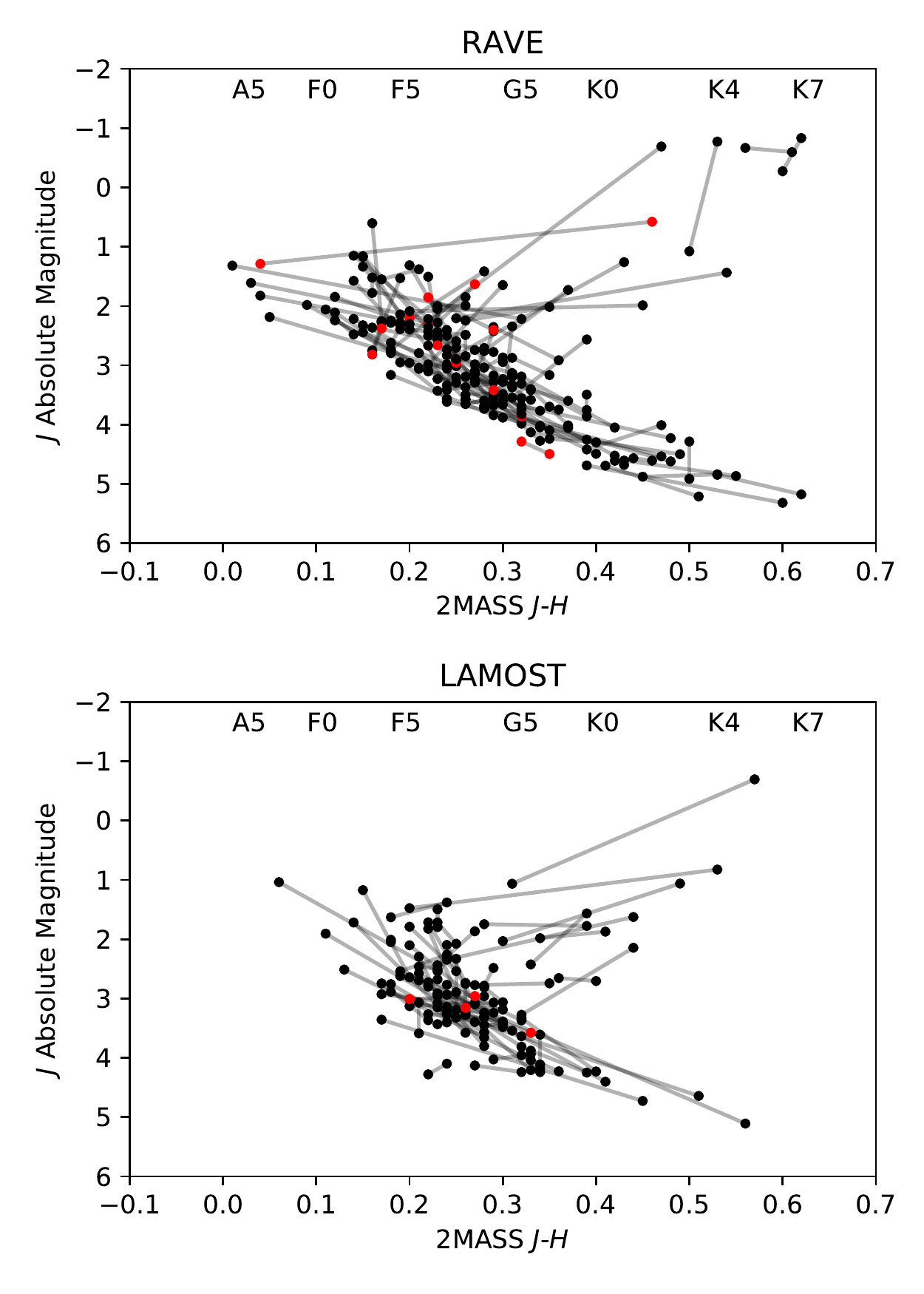}
    \caption{The colour-magnitude diagram of wide binaries in our RAVE (top panel) and LAMOST samples (bottom panel). At the top of each panel, we provide spectral types corresponding to $J$-$H$ colours as listed by \citet{kraus07}. No correction for reddening is included. Not only do the components in our pairs span a range of stellar types, but lines linking the two components of each pair also show that pairs are often comprised of differing stellar types. Furthermore, although the majority of the stars in our candidate binaries are on the main sequence, several of our pairs include evolved stars. Pairs with metallicities consistent (inconsistent) at the 3$\sigma$ confidence level are denoted in black (red). In Section~\ref{sec:delta_metallicity_source} we provide a few potential mechanisms that may account for these metallicity differences, including mixing processes occurring near and after the main sequence turn-off.}
    \label{fig:HR}
\end{figure}

We also cross-match with the LAMOST DR3 catalog to identify an additional 85 wide binaries with measured RVs, although only 62 of these have measured metallicities with uncertainties less than 0.2 dex. These 115 RAVE binaries and 62 LAMOST binaries, which have full six-dimensional phase space measurements and metallicities, form the sample of wide binaries we use throughout this work\footnote{Data driven metallicity and abundance catalogs, such as the RAVE-on catalog \citep{casey17} and the catalog of LAMOST giants by \citet{ho17} offer potentially more precise measurements for a larger range of elements. Unfortunately, these catalogs focus on giant stars and have little overlap with our sample of TGAS wide binaries.}. As in Table \ref{tab:RAVE}, the {\it Gaia} Source IDs for the stars in these 62 pairs are provided in Table \ref{tab:LAMOST} along with their RVs, metallicities, surface gravities and effective temperatures, as measured by LAMOST.

Figure \ref{fig:HR} shows a colour-magnitude diagram of the stars in the RAVE (top panel) and LAMOST (bottom panel) samples of binaries generated by combining {\it Gaia} parallaxes with Two Micron All Sky Survey \citep[2MASS;][]{2mass} photometry. Lines link the two components of each wide binary. Labels at the top of each panel indicate the 2MASS $J$-$H$ colours corresponding to different stellar types \citep[as defined by][]{kraus07}; the stars in our sample span a large range from early K-type to late A-type stars, and several have evolved off the main sequence.

\begin{figure}
	\includegraphics[width=0.9\columnwidth]{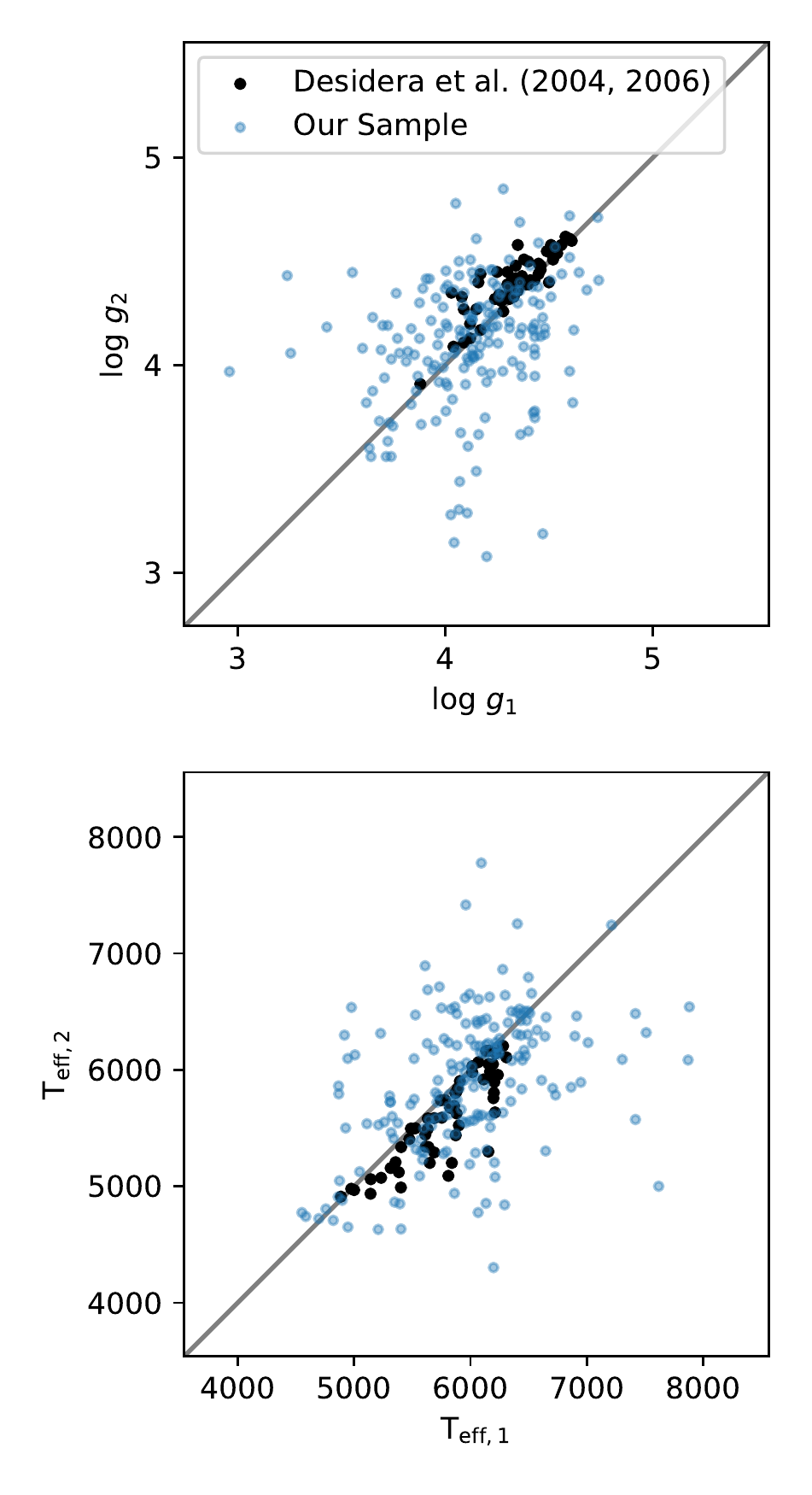}
    \caption{The log $g$ (top panel) and T$_{\rm eff}$ (bottom panel) of the components of wide binaries both in our sample (blue) and the sample from \citet[black;][]{desidera04,desidera06b}. Our sample more than triples the number of wide binaries with spectroscopically measured properties, and the components in our pairs span a wider range in log $g$ and T$_{\rm eff}$. }
    \label{fig:logg_teff}
\end{figure}

In Figure \ref{fig:logg_teff}, we compare the log $g$ and T$_{\rm eff}$ for the stars in the binaries in our sample (blue points) with the sample of 54 binaries by \citet[black points;][]{desidera04,desidera06b}, the largest to date in which metallicities have been studied. Previous samples have principally focused on pairs of solar-type stars with similar T$_{\rm eff}$ and log $g$. Not only is our sample larger in number, but it includes stars of very different types.

\begin{figure}
	\includegraphics[width=1.03\columnwidth]{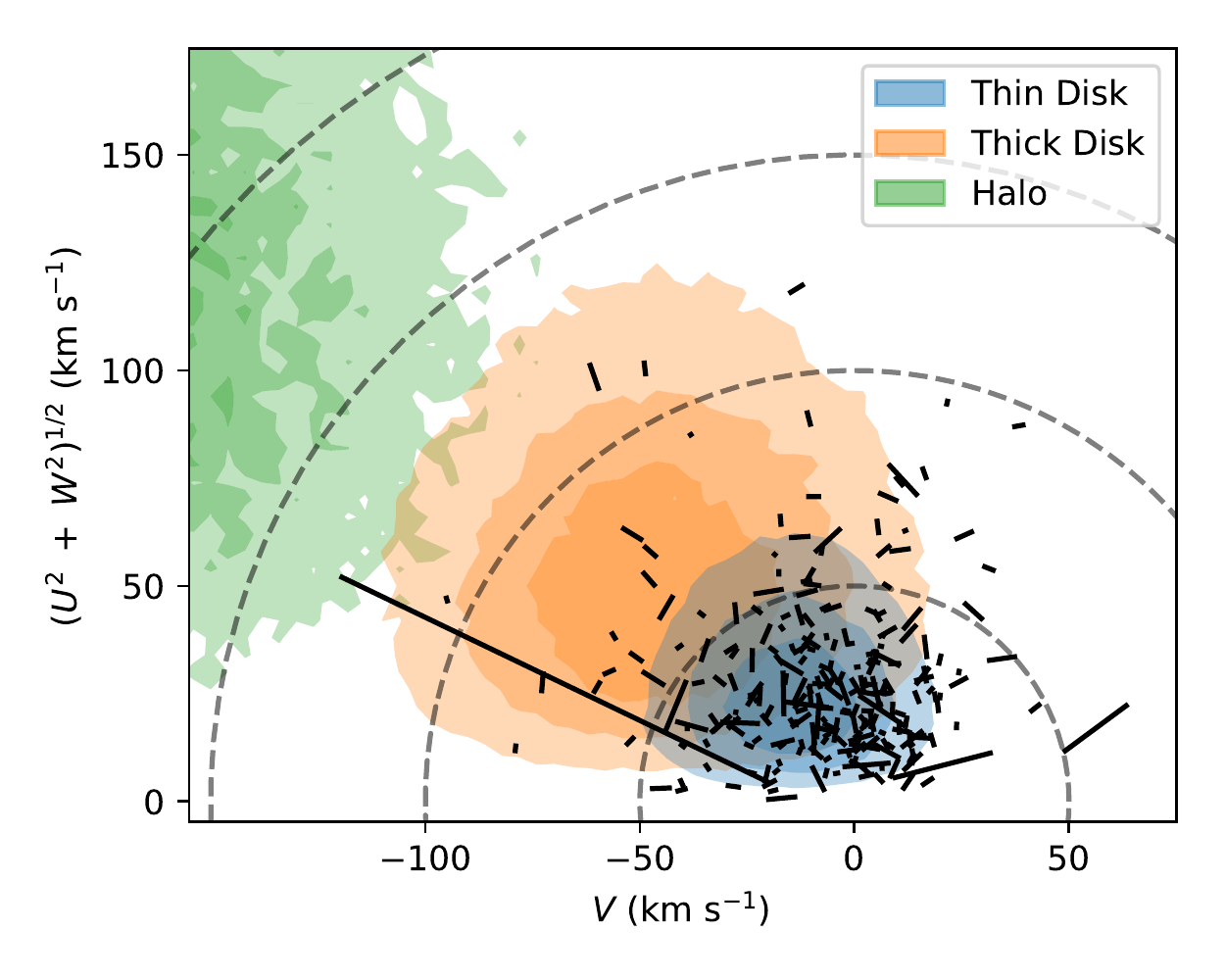}
    \caption{Toomre diagram of the TGAS wide binaries. Black lines link the two stars in each wide binary pair. Background colours correspond to the kinematic thin disk (blue), thick disk (orange), and halo (green). The majority of pairs are associated with the thin disk, but several pairs are likely to be thick disk members. None of our pairs are clear halo binaries. Longer lines indicate binaries with either poorly constrained or inconsistent RVs. }
    \label{fig:toomre}
\end{figure}

We also use {\it Gaia} astrometry and RAVE and LAMOST RVs to associate the wide binaries in this sample with kinematic components of the Milky Way. We use the Milky Way population model of \citet{bensby03} to convert phase space measurements into (right-handed) $U$, $V$, and $W$ velocities. The results of this conversion are shown in the Toomre diagram\footnote{A Toomre diagram shows the relationship between the rotational velocity and the vertical and radial velocities of stellar orbits in the Galaxy \citep[see discussion about its use for studies of Galactic structure in e.g.,][]{Qu11}.} in Figure~\ref{fig:toomre}. Black lines join the components of each pair in our sample; longer lines indicate systems that have either discrepant or poorly measured RVs. Our sample is predominantly comprised of kinematically thin disk stars, although there may be a number of thick disk pairs as well. There are no binaries in our sample that are clearly associated with the halo. We therefore expect these pairs to have metallicities similar to solar.

\section{The Relative Metallicities of Our Wide Binaries}
\label{sec:metallicity}

\begin{table}
\begin{center}
\caption{$p$-values for correlation tests between the metallicities and elemental abundances of wide binary components. Corresponding correlation coefficients are provided in each panel in Figure \ref{fig:metallicity}, \ref{fig:abundance_genuine}, and \ref{fig:abundance_false}. \label{tab:p_values} }
\begin{tabular}{lcc}
\toprule
 & Wide Binaries & Random Alignments \\
\midrule
\multicolumn{3}{c}{LAMOST Metallicity} \\
\midrule
Pearson & 1.80$\times 10^{-14}$ & 0.46 \\
Spearman & 2.27$\times 10^{-14}$ & 0.33 \\
Kendall & 7.70$\times 10^{-14}$ & 0.35 \\
\bottomrule
\multicolumn{3}{c}{RAVE Metallicity} \\
\midrule
Pearson & 8.11$\times 10^{-7}$ & 0.88 \\
Spearman & 1.29$\times 10^{-6}$ & 0.67 \\
Kendall & 8.43$\times 10^{-7}$ & 0.64 \\
\bottomrule
\multicolumn{3}{c}{RAVE [Fe/H]} \\
\midrule
Pearson & 1.40$\times 10^{-10}$ & 0.11 \\
Spearman & 3.61$\times 10^{-11}$ & 0.15 \\
Kendall & 2.89$\times 10^{-10}$ & 0.12 \\
\bottomrule
\multicolumn{3}{c}{RAVE [Mg/H]} \\
\midrule
Pearson & 6.32$\times 10^{-5}$ & 0.28 \\
Spearman & 2.23$\times 10^{-5}$ & 0.37 \\
Kendall & 3.98$\times 10^{-5}$ & 0.41 \\
\bottomrule
\multicolumn{3}{c}{RAVE [Al/H]} \\
\midrule
Pearson & 2.63$\times 10^{-4}$ & 0.47 \\
Spearman & 8.63$\times 10^{-4}$ & 0.62 \\
Kendall & 1.06$\times 10^{-3}$ & 0.63 \\
\bottomrule
\multicolumn{3}{c}{RAVE [Si/H]} \\
\midrule
Pearson & 3.75$\times 10^{-10}$ & 0.80 \\
Spearman & 1.34$\times 10^{-11}$ & 0.87 \\
Kendall & 1.56$\times 10^{-10}$ & 0.78 \\
\bottomrule
\multicolumn{3}{c}{RAVE [Ti/H]} \\
\midrule
Pearson & 3.89$\times 10^{-2}$ & 0.22 \\
Spearman & 1.12$\times 10^{-2}$ & 0.46 \\
Kendall & 8.52$\times 10^{-3}$ & 0.39 \\
\bottomrule
\end{tabular}
\end{center}
\end{table}

\begin{figure*}
	\includegraphics[width=0.9\textwidth]{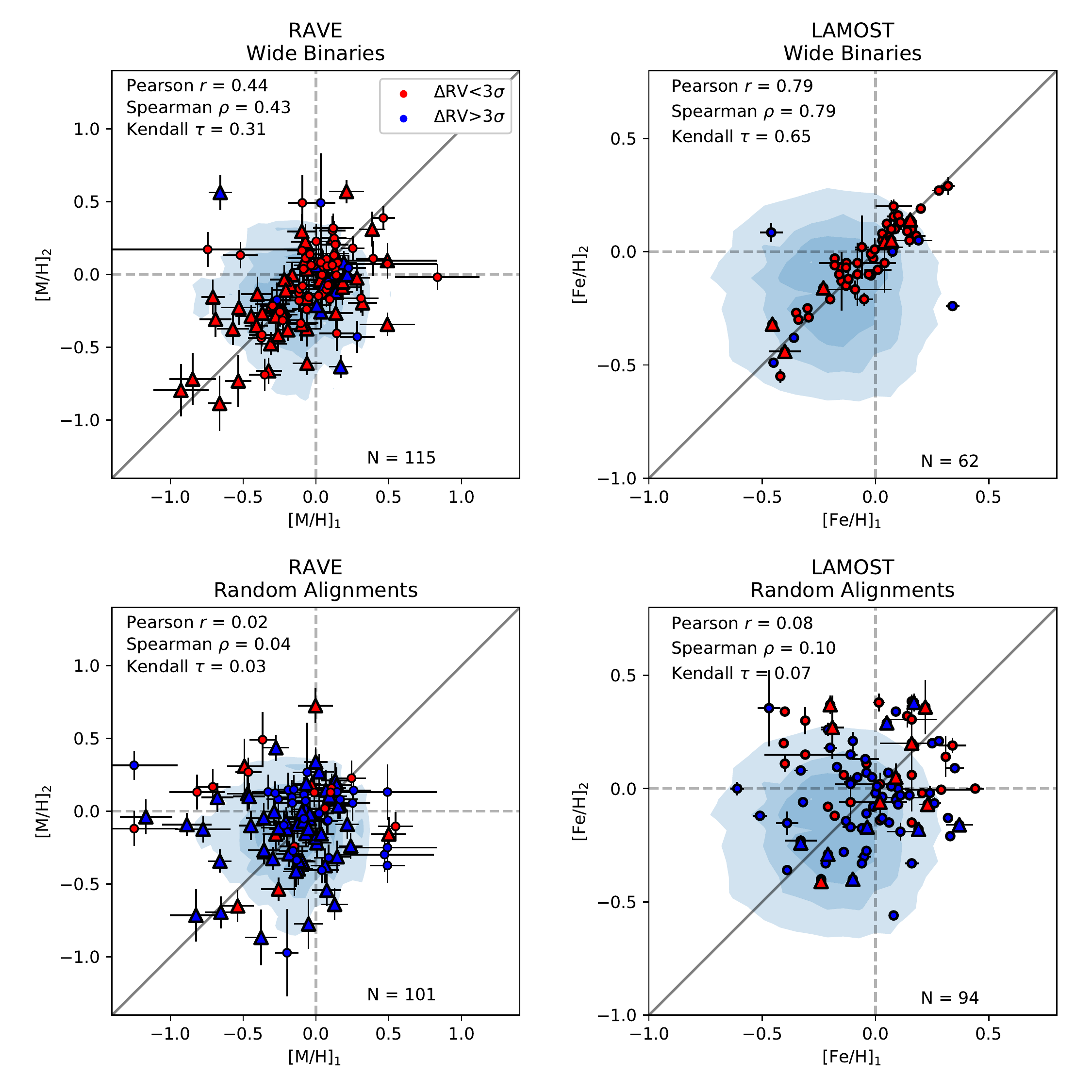}
    \caption{The top panels compare the metallicities of the two components of each candidate wide binary in our catalog using either RAVE (left panels) or LAMOST data (right panels). Note that we follow each study's separate metallicity notation. Blue data points indicate binaries with RVs inconsistent at the 3-$\sigma$ level. Pairs with at least one star with log $g$ $<3.8$ are denoted by triangles and outlined in black. The blue background indicates the expectation from stars randomly paired without regard to their kinematics or parallaxes. The bottom panels shows the corresponding metallicity comparison for pairs generated from random alignments. Correlation coefficients in the top left of each panel show quantitatively that our sample of genuine pairs is positively correlated ($p$-values are less than 10$^{-6}$ for all three correlation statistics), while the sample of random alignments in the bottom panels have uncorrelated metallicities, as expected. Despite a few outliers from the one-to-one line, the components of our sample of wide binaries show a statistically significant correlation.}
    \label{fig:metallicity}
\end{figure*}

Figure \ref{fig:metallicity} compares the metallicities of the two components of each binary in our sample for both those pairs in RAVE (top left panel) and LAMOST (top right panel).\footnote{\label{fn:metals} Although both RAVE and LAMOST identify an overall stellar metallicity from their spectra based on automated pipelines, RAVE denotes this quantity as [M/H] \citep{kordopatis11} while LAMOST uses [Fe/H] \citep{wu14}. We elect to retain each catalog's separate notation.} We separate these binaries into those with consistent RVs (red; $\Delta$RV$<3\sigma$), those with inconsistent RVs (blue; $\Delta$RV$>3\sigma$). Pairs in which at least one star is a giant \citep[log $g$ $<$ 3.8;][]{zwitter10} have triangular symbols. Blue backgrounds are generated by taking pairs of random stars from either the RAVE or LAMOST catalogs.

The metallicities of components of our wide binaries cluster around the one-to-one line, although clear outliers exist. We provide correlation coefficients for three different statistical correlation tests in the top left of each panel, and the corresponding $p$-values in Table~\ref{tab:p_values}. These are calculated from the subset of pairs with consistent RVs. These tests all provide $p$-values $<$10$^{-6}$, indicating that a significant correlation exists.

Binaries measured by LAMOST tightly correlate around the one-to-one line. Although perhaps not as striking, the binaries measured by RAVE still show a clear and significant correlation as measured by the correlation coefficients and their corresponding $p$-values (provided in Table \ref{tab:p_values}). The differences between the two data sets derive principally from the improved quality of LAMOST DR3 metallicity measurements. LAMOST provides metallicity measurements with formal uncertainties $<$0.01 dex for a large fraction of stars in our sample. Of the 62 binaries measured by LAMOST, the two with obviously discrepant metallicities also have significantly differing RVs, indicating potential contamination, or the presence of unseen companions to one of the wide binary components. The source of metallicity discrepancies in binaries measured by RAVE is much less clear because of the larger measurement uncertainties. We discuss this further in Section \ref{sec:discussion}.

For a null hypothesis comparison sample, we use the set of randomly aligned stellar pairs produced in Paper I. This sample was generated using the identical method we used to identify our sample of wide binaries, except we match stars in the TGAS catalog to a version of itself in which the positions and velocities of the stars are shifted; every stellar pair in this catalog is, by design, a random alignment. To obtain a large enough sample for comparison, we select pairs without any limit on the projected separation and reduce the restriction on the posterior probability assigned to the pair by our model from 99\% (as used in our sample of wide binaries) to 50\%. Despite loosening the posterior probability constraint, these pairs still have closely matching proper motions and parallaxes. From this, we obtain a sample of 101 randomly aligned stellar pairs with RAVE, and 94 with LAMOST, measurements.

The bottom panels of Figure \ref{fig:metallicity} show the metallicity comparison for pairs from our sample of random alignments. The difference between the bottom and top panels for each data set is stark. Correlation tests give $p$-values for these samples of 0.3-0.9 (provided in Table \ref{tab:p_values}), quantitatively indicating what is obvious from a glance: randomly aligned stars show no evidence of metallicity correlation. Although this fact may at first seem unsurprising, for two stars to be selected as a random alignment, they must have similar proper motions and parallaxes (despite the shifts in the catalog) and therefore be part of the same broadly defined kinematic component of the Galaxy. Since they are from the same kinematic component of the Galaxy, one might expect these random alignments to show some metallicity correlation, yet we see no evidence for one. We discuss the implications of this further in Section \ref{sec:discussion}. Note also that the RVs of these pairs are largely inconsistent, (i.e., a majority of blue points in comparison to the upper panels, where the majority are red) as expected from randomly aligned stars.

\begin{figure}
	\includegraphics[width=\columnwidth]{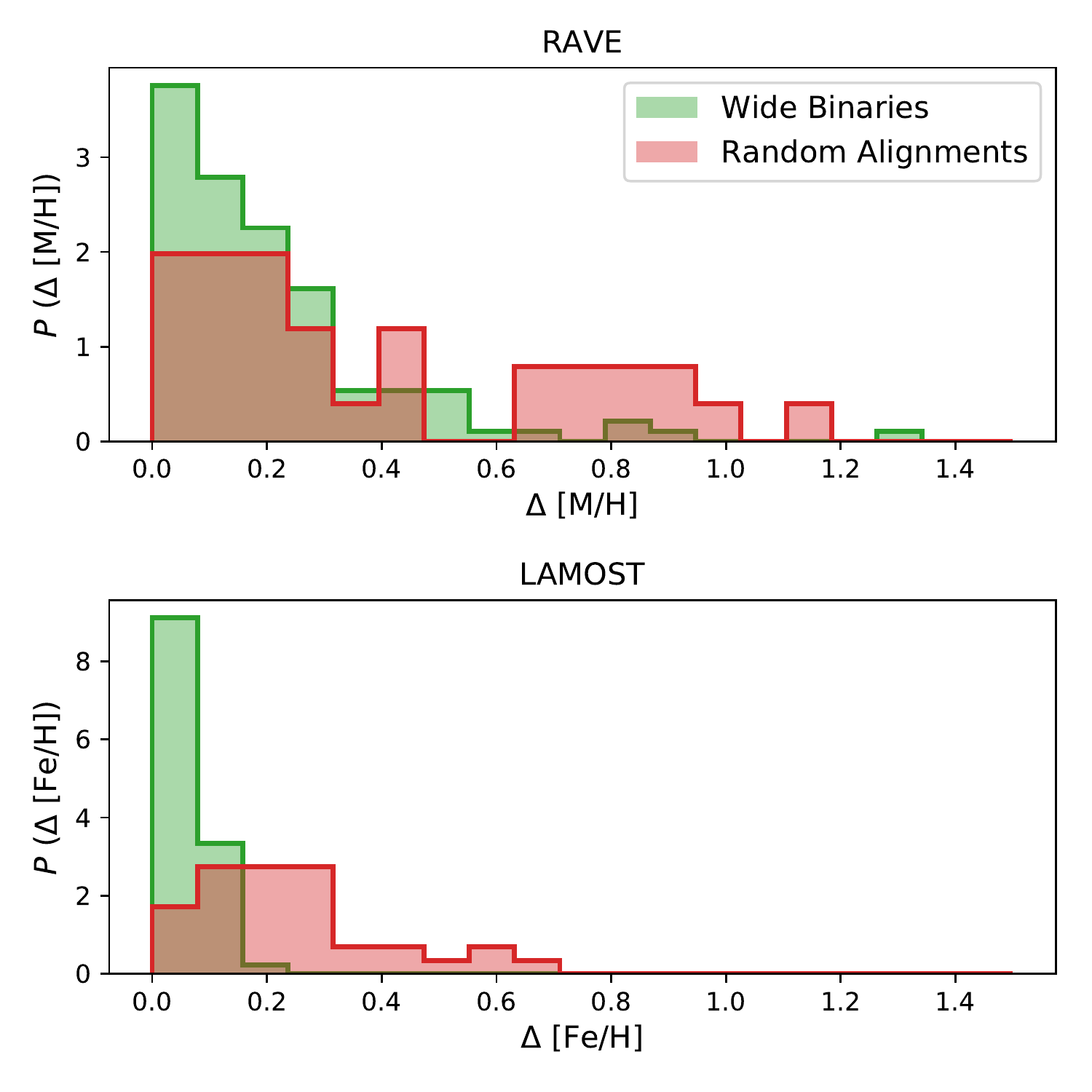}
    \caption{Normalized distributions of the metallicity differences of our RAVE (top panel) and our LAMOST samples (bottom panel) between our catalog of genuine pairs (green) and random alignments (red). }
    \label{fig:delta_metallicity}
\end{figure}

\begin{figure}
	\includegraphics[width=\columnwidth]{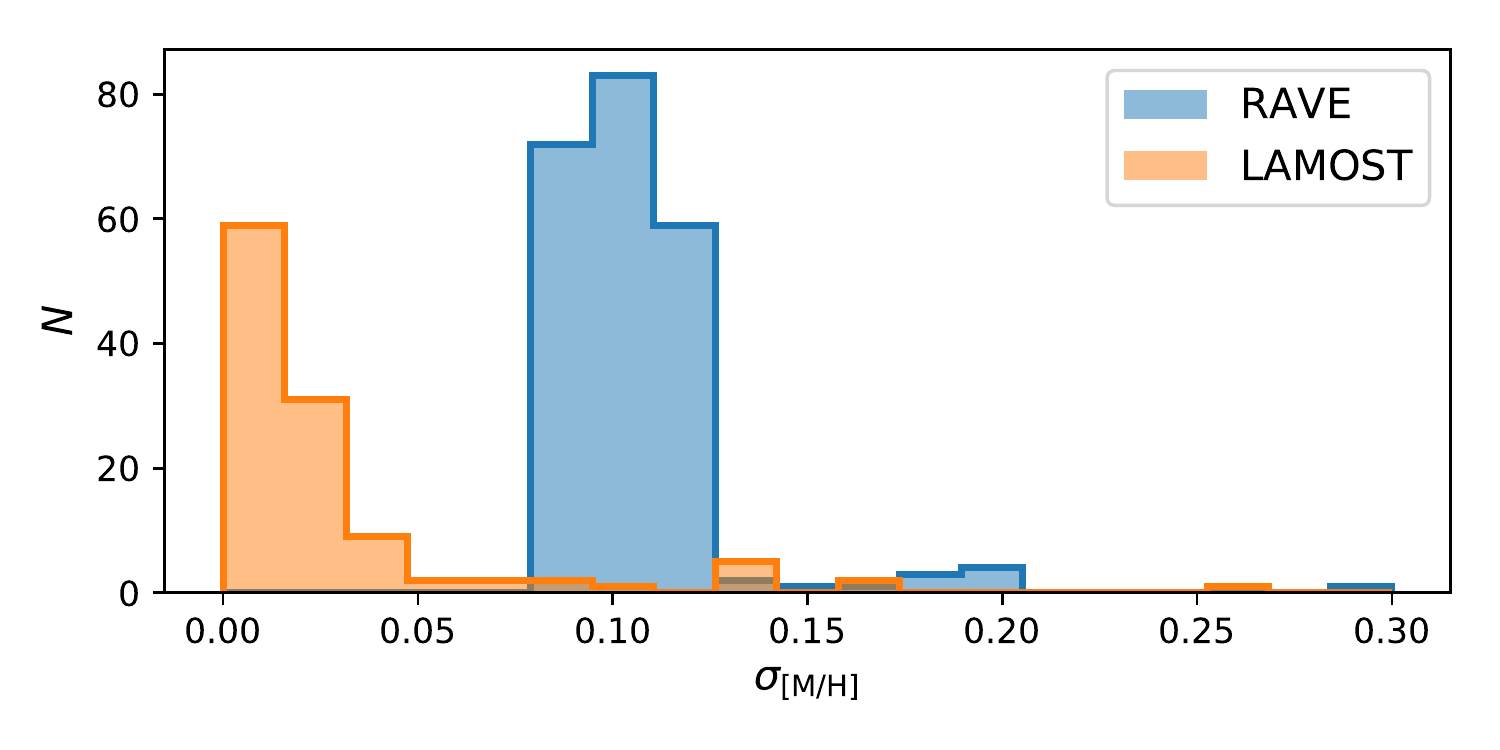}
    \caption{As reported by the surveys, LAMOST DR3 metallicities are more precise by nearly an order of magnitude than RAVE metallicities. The extreme precision of the LAMOST metallicity uncertainties derive from the high signal-noise ratios for these spectra (often in excess of 200) since the TGAS stars in these pairs are relatively bright ($V\lesssim$12). }
    \label{fig:M_H_uncertainty}
\end{figure}

As a further test, we compare the distribution of metallicity differences in Figure \ref{fig:delta_metallicity} for both our wide binary sample (green) and random alignment sample (red) for our sample with RAVE and with LAMOST measurements. Our wide binary sample clusters around $\Delta$[Fe/H]~$=0$; however, outliers with large metallicity differences exist. Figure \ref{fig:M_H_uncertainty} shows the distribution of [Fe/H] uncertainties in the RAVE and LAMOST samples. The nearly order-of-magnitude smaller uncertainties in the LAMOST measurements allow us to see that the distribution of $\Delta$[Fe/H] is closely peaked at 0 in the bottom panel of Figure \ref{fig:delta_metallicity} which accounts for the tight correlation in the top right panel in Figure \ref{fig:metallicity}. Because TGAS stars are relatively bright ($V\lesssim$12), typical LAMOST spectra of our sample have very high S/N, often in excess of 200. Therefore, LAMOST metallicities are provided with precisions of order a few 0.01 dex. We discuss this further in Section \ref{sec:discussion_abundances}.

\begin{figure*}
	\includegraphics[width=\textwidth]{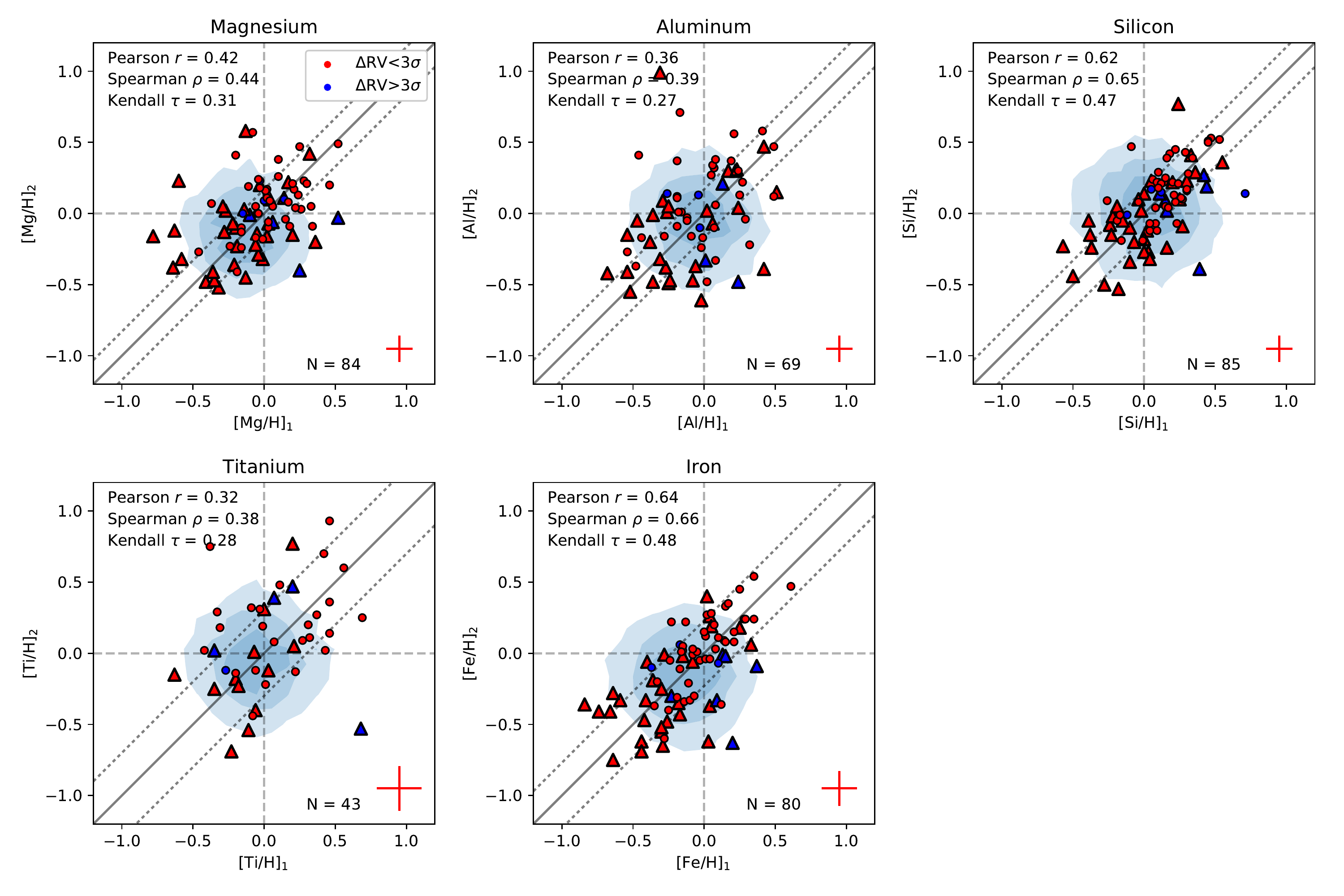}
    \caption{Abundance comparison of the components of genuine pairs for five different elements ([Fe/H] is calculated here from Fe lines). Triangles denote those pairs in which at least one component has a log $g$ $<3.8$. The sample sizes for each element differ since only a subset of all stars observed by RAVE have measurable abundances in each element. Blue backgrounds indicate the expectation of RAVE stars randomly paired together without regard to their kinematics or parallaxes. One-to-one lines along with the $+1\sigma$ and $-1\sigma$ (dashed) are shown in each panel. Typical uncertainties on the abundance measurements of stars in the catalog are indicated by the red cross at the bottom right of each panel. Correlation coefficients provided in the top left of each panel show that these data are significantly correlated; $p$-values are provided in Table \ref{tab:p_values} and quantitatively show significant correlations for all tested elements but Ti.}
    \label{fig:abundance_genuine}
\end{figure*}

\begin{figure*}
	\includegraphics[width=\textwidth]{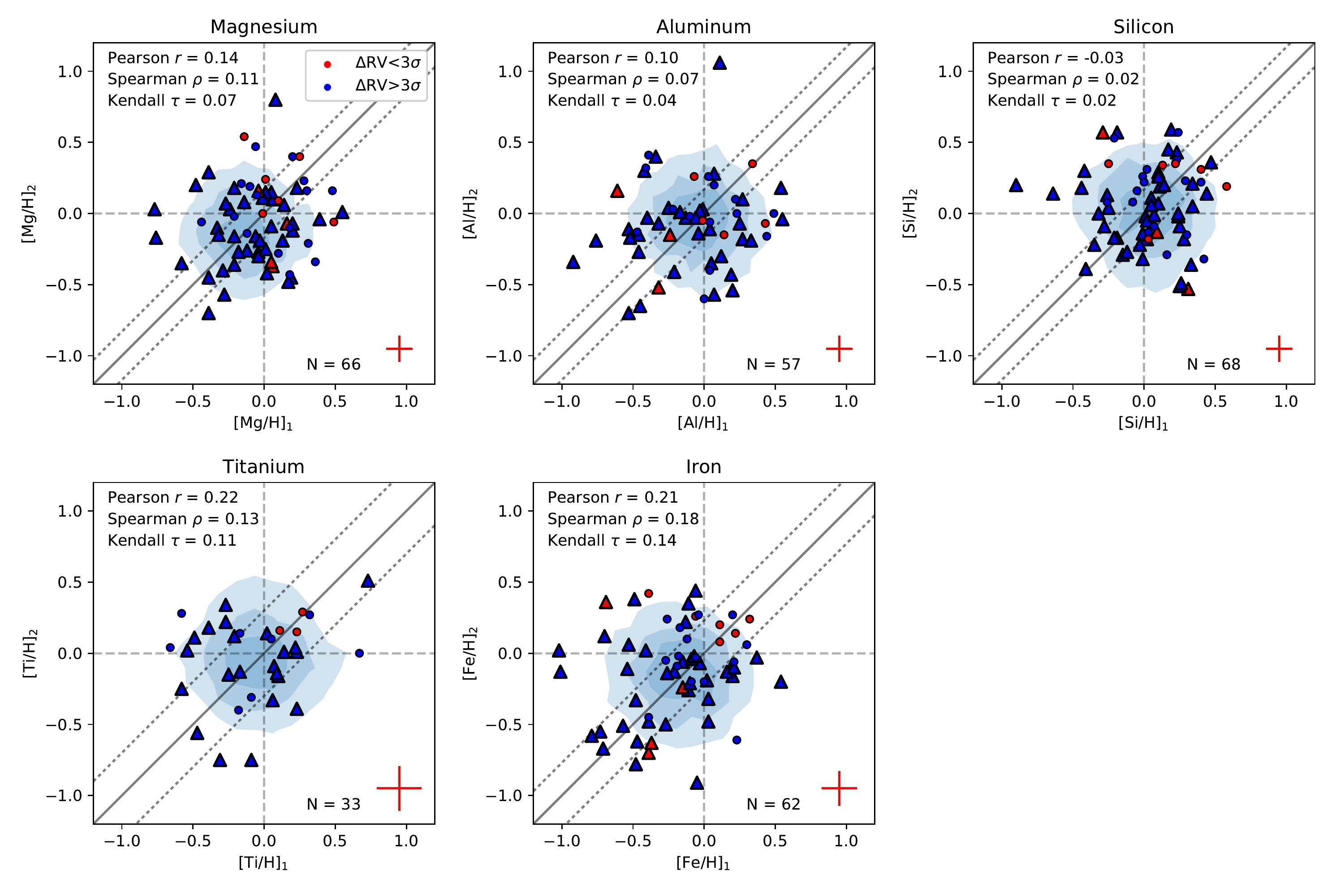}
    \caption{Abundance comparison of the components of random alignments. Plot symbols, backgrounds, and one-to-one lines are all identical to Figure \ref{fig:abundance_genuine}. Correlation coefficients quantitatively indicate that the two stars in each of these pairs are uncorrelated. $p$-values are provided in Table \ref{tab:p_values}.}
    \label{fig:abundance_false}
\end{figure*}

\section{The Relative Elemental Abundances of Our Wide Binaries}
\label{sec:abundance}

In addition to metallicities, calculated from the Ca triplet, RAVE also supplies the abundances of a number of elements, including Fe, as calculated directly from metallic lines. Figure \ref{fig:abundance_genuine} compares the abundances of the components of each of our pairs for Mg, Al, Si, Ti, and Fe.\footnote{Where possible, RAVE additionally measures the Ni abundance; however, too few pairs have measurements for both components to allow a critical analysis of Ni.} Each panel contains a variable number of pairs since only a subset of all RAVE spectra have measurable lines for each of these elements. The red cross at the bottom right in each panel indicates the uncertainty of each elemental abundance measurement which translates directly into 1$\sigma$ uncertainties (dashed line) around the one-to-one line (solid line) between the two stars' abundances. As in Figure \ref{fig:metallicity}, the blue background is generated by randomly pairing RAVE stars without regard to their kinematics or parallaxes.

With few exceptions, the stars in our pairs share consistent abundances. Visually, stars with sub- or super-solar abundances tend to have companions with sub- or super-solar abundances. Quantitatively, correlation coefficients at the top left of each panel show positive correlations ($p$-values for our correlation tests are provided in Table \ref{tab:p_values} and show statistically significant correlations). It is worth noting that the [Fe/H] abundance as measured by RAVE has a stronger correlation than the overall RAVE metallicity. As with our metallicity comparison in Figure~\ref{fig:metallicity}, a subset of our binaries have substantially different abundances, potentially indicating contamination. We discuss these pairs further in Section \ref{sec:discussion_abundances}.

In Figure \ref{fig:abundance_false}, we compare the abundances of the components of random alignments. The panels are analogous to those in Figure \ref{fig:abundance_genuine}. Correlation coefficients and their corresponding $p$ values (shown in Table \ref{tab:p_values}) indicate that these pairs have near-zero correlation, and indeed they seem to overlap significantly with the blue backgrounds indicating randomly paired RAVE stars.

\begin{figure}
	\includegraphics[width=\columnwidth]{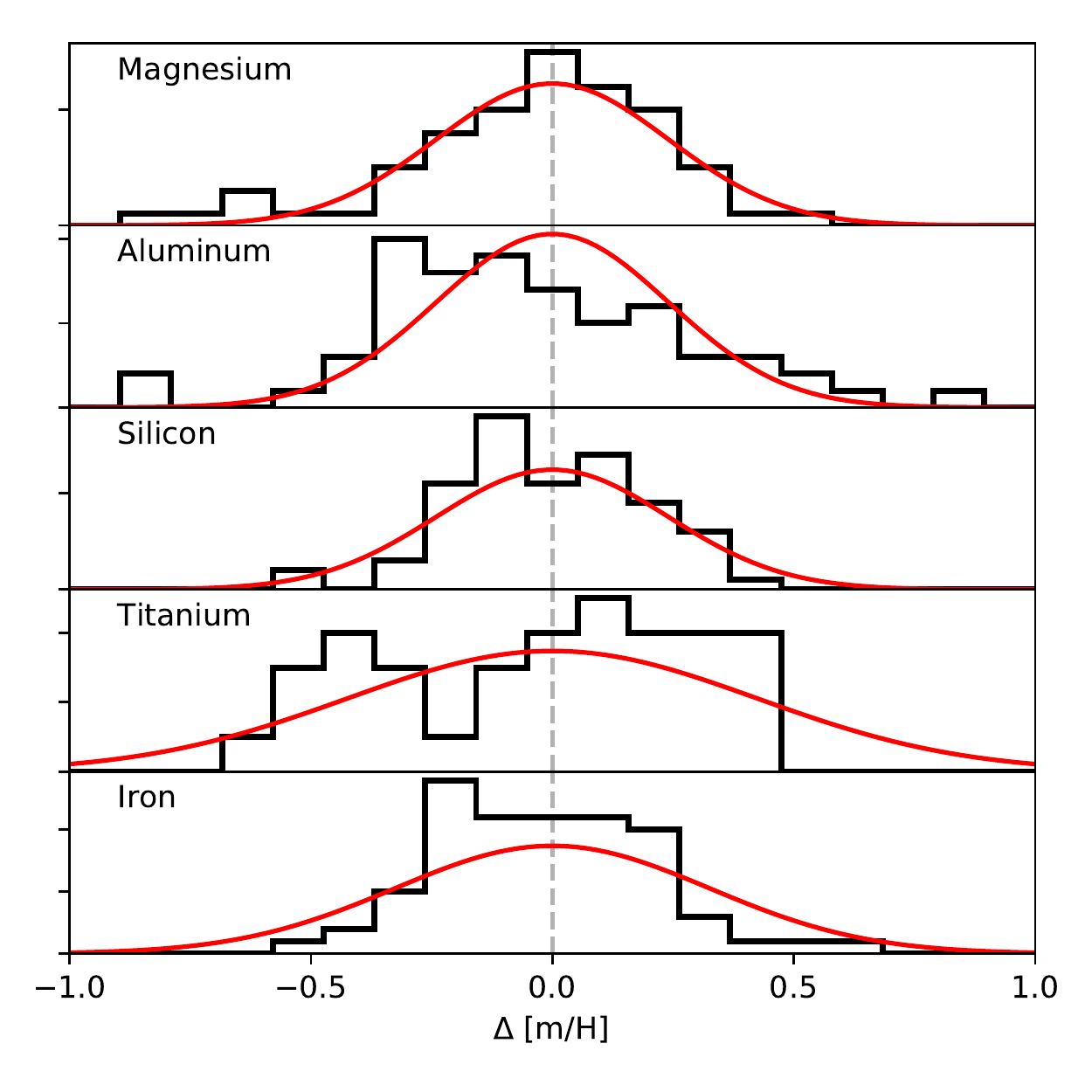}
    \caption{Differences in the abundances (black) of the components of genuine pairs for six different elements. Gaussian curves (red) indicate the expected distribution based on RAVE uncertainties for each element. The abundances are consistent with the measurement precision. }
    \label{fig:delta_abundance}
\end{figure}

\section{Discussion}
\label{sec:discussion}

\subsection{The components of wide binaries have the same metallicity}
\label{sec:discussion_abundances}

Our study builds off previous work by \citet{desidera04,desidera06b}, who showed that [Fe/H] abundances of the components of 56 wide binaries typically differ by no more than 0.02 dex, and all but one pair in their sample have $\Delta$[Fe/H] $<$ 0.1 dex. The singular outlier (HIP 64030) has an anomalously large metallicity difference ($\Delta$[Fe/H] = 0.27 dex), and may host a blue straggler \citep{desidera07}.

Although the top panel of Figure~\ref{fig:delta_metallicity} shows that many binaries in our sample have $\Delta$[Fe/H] $>$ 0.1 dex as measured by RAVE, RAVE metallicity measurements have relatively larger uncertainties. The distribution of $\Delta$[Fe/H] has a standard deviation of 0.29 dex, somewhat larger than what is expected from observational uncertainties in the [Fe/H] measurements, which are typically of order 0.1 dex (See Figure~\ref{fig:M_H_uncertainty}). A small amount of contamination and samples with large uncertainties may skew the distribution of $\Delta$[Fe/H] toward larger values.

However, the more precise LAMOST metallicity measurements show a particularly clear correlation, which indicates that the principal source of metallicity differences in the RAVE sample are due to observational uncertainties. LAMOST metallicity differences are typically less than 0.1~dex, as shown by the green histogram in the bottom panel of Figure \ref{fig:delta_metallicity}, with the largest difference being 0.16 dex. Formally, while the uncertainties in individual stars' metallicities may be listed as $<$0.01 dex, systematic, repeat measurements of the same stars by the LAMOST team show an internal uncertainty of $\approx$0.06 dex \citep{luo15}, which would correspond to a $\Delta$[Fe/H] uncertainty of 0.09 dex. This uncertainty likely varies with magnitude, and as our sample is systematically brighter than typical LAMOST targets, the internal uncertainties for our sample are likely somewhat smaller. Indeed, the distribution of $\Delta$[Fe/H] for our LAMOST sample has a standard deviation of 0.07 dex, smaller than the expected 0.09 dex.

With our sample, we can therefore independently confirm the conclusions of \citet{gratton01} and \citet{desidera04,desidera06b} that the components of wide binaries have consistent metallicities. The differences we see are consistent with being due to observational uncertainties.

The data from the RAVE catalog allow us to extend the chemical comparison beyond an overall metallicity to abundances of Mg, Al, Si, Ti, and Fe. Here again, the $p$-values calculated from our statistical tests indicate that there is a significant, positive correlation between the abundances of these elements, whereas the random alignments show no correlation. Figure \ref{fig:delta_abundance} indicates that the differences in the distribution of abundances roughly corresponds to the precision of RAVE measurements. Similar to [Fe/H], the components of wide binaries have consistent abundances. Higher precision measurements are required to determine the level of consistency, but with only a few exceptions, abundance differences are consistent with being due to observational uncertainties.

\subsection{What are the implications for wide binary formation scenarios?}
\label{sec:formation_scenarios}

As the LAMOST data in particular indicate, wide binaries have metallicities substantially more consistent than pairs of random field stars. This is consistent with canonical formation scenarios involving the disruption of a low-mass stellar cluster \citep[outlined by][]{kouwenhoven10,moeckel11} and the unfolding of a hierarchical triple system \citep[outlined by][]{reipurth12}. The recent suggestion by \citet{tokovinin17} that wide binaries may form from neighbouring pre-stellar cores would also seem to imply that wide binaries should have nearly identical metallicities.

Dynamical formation scenarios, on the other hand, would form wide binaries from random field stars, producing binaries with uncorrelated metallicities. Such scenarios are strongly disfavored on theoretical arguments \citep{mansbach70,bodenheimer92}. Were it a relevant mechanism, one might expect that the most likely site for wide binary formation would be in kinematically similar stellar structures such as moving groups. \citet{bovy10} show that moving groups have metallicities consistent with background stars from the Milky Way disk. Taken together with our results, the implication is clear: wide binaries cannot be formed in substantial numbers within moving groups. Therefore, the recently discovered wide binaries in the $\beta$-Pictoris moving group by \citet{elliott16} are likely to be primordial, in agreement with these authors' conclusions.

The level of general metallicity and specific elemental consistency between components of wide binaries implies that the large majority of wide binaries are likely to be genuinely co-chemical and co-eval. Although more precise measurements are needed, elemental abundances measured by RAVE also show a significant correlation that cannot be accounted for by stars with different initial chemistry. Therefore, our sample confirms that wide binaries may be safely used as calibrators for age-activity-rotation relations, the initial-final mass relation for white dwarfs as well as M-dwarf metallicity indicators.

\subsection{What are the source of wide binary metallicity differences?}
\label{sec:delta_metallicity_source}

Figure \ref{fig:metallicity} shows the differences in the metallicities of wide binary components is consistent with being due to observational uncertainties. The LAMOST data in the top right panel of this figure presents the most striking demonstration: the two pairs with components showing significantly different metallicities also have inconsistent RVs, indicating these are likely due to random alignments. Two random alignments out of the 62 in our sample is roughly consistent with our estimate from Paper I of a 5-10\% contamination rate. From these data, we can say the level of metallicity consistency is $<$0.1 dex, which can be used to gauge the effect of various physics expected to affect surface abundance measurements.

For instance, \citet{desidera06b} demonstrate that, although most wide binaries have differences $<$0.02 dex in [Fe/H], if the stars in a given pair have larger differences in temperature, the metallicity difference as observed in stellar spectra can increase. This may be due to non-LTE effects such as overexcitation; \citet{schuler10} show that the iron and oxygen abundances of stars in the Pleiades differ depending on which ionization state is used to derive the abundances. This effect is likely to cause differences of the order of 0.05 dex (depending on the stellar temperature/mass) and may account for some of the scatter in the metallicities.

Due to standard changes after first dredge-up, and likely involving some sort of non-standard extra mixing as well, the abundances of light elements in the surfaces of red giants are well known to diverge from those observed in dwarfs from the same parent population \citep[e.g.,][]{kraft94,pasquini04,chaname05, schuler09}. \citet{dotter17} recently quantified the effect of atomic diffusion and mixing processes on stellar abundances, concluding that these may affect the metallicities of stars at a level of 0.1-0.2 dex, with the largest differences occurring at the main sequence turn-off and the giant branches. These authors suggest that any effects would be mitigated by selecting stars at similar evolutionary states (though, of course, evolutionary state is a rather hard thing to determine reliably for field stars).

Red points in the CMD in Figure~\ref{fig:HR} indicate those stars with discrepant metallicities. Some of the pairs with discrepant metallicities indeed include giant stars or stars near the main sequence turn-off. Evolutionary differences may be responsible for some of the observed metallicity differences, particularly in the RAVE sample.

Alternatively, abundance differences could be due to accretion by planetary material \citep{gonzalez97}. Hot Jupiters have the potential to scatter smaller, rocky bodies into their host star, altering their abundances \citep{ida08}. For instance, \citet{mack14} argue that the abundance difference of $\approx$0.04 dex between the components of the planet-hosting wide binary HD 20782/81 could be due to ingestion of 10-20 $M_{\oplus}$ of rocky, planetary material. As a further example, \citet{oh17b} identify the wide binary HD 240429/30 as being comprised of two G stars with an abundance difference of $\approx$0.2 dex in refractory elements which these authors attribute to engulfment of 15 $M_{\oplus}$ of rocky material. More precise abundance measurements, particularly of the refractory elements, are required to test this hypothesis, but it does allow for the possibility that wide binaries with different metallicities may be an indicator of the presence of planets \citep{desidera04, desidera06b}.

Rather than comets or small planetary debris, engulfment of low-mass of sub-stellar companions can cause substantial elemental abundance differences, as modeled by \citet{siess99a, siess99b}. For instance, \citet{aguilera16b} recently argued that engulfment may be the cause of a class of Li-rich red giants in Trumpler 20.

\subsection{Can we use abundances to identify candidate wide binaries?}
\label{sec:abundances_for_matching}

Previous studies of small samples and of individual wide binaries \citep[e.g.,][]{desidera06b,liu14}, as well as the results presented here, indicate that the components of wide binaries should have similar (possibly near-identical) chemistry. This presents the intriguing possibility that abundance distributions could be used as criteria for identifying wide binaries in the same way as it is being used to identify stellar structures within the Milky Way \citep{freeman02}. Modern spectroscopic surveys with individual elemental abundance measurements for large numbers of stars are quickly becoming available.

The bottom left panel of Figure \ref{fig:metallicity} demonstrates that metallicity alone, at the precision of RAVE, is not a good indicator of a common origin since most ($\approx$80\%) random alignments have metallicities consistent at 3$\sigma$. When comparing stars with more precise LAMOST DR3 measurements, the fraction of random alignments with metallicities consistent at 3$\sigma$ drops to 34\%. In tandem with precise astrometry, metallicities measured at a precision of a few 0.01 dex can likely help separate genuine binaries from contamination.

Matching abundances of several elements simultaneously may be an even more effective method to associate specific stars. However, we find that of our 101 randomly aligned stellar pairs with RAVE elemental abundances, only 13 have a discrepancy in the abundance of at least one element at the 3$\sigma$ level. Using a stricter constraint of 1$\sigma$, slightly less than half of all random alignments have consistent abundances. Therefore, abundance measurements more precise than those available in RAVE are required for elemental abundances to be included as a selection criteria for wide binaries.

At the same time, some {\it bona fide} wide binaries may have significantly different abundances, and these pairs may be the most interesting for follow-up study. For instance, \citet{desidera07} showed that the wide binary HD 113984, whose components have substantially different abundances, is a unique triple system with a blue-straggler component. Of course, such systems are intrinsically rare. We note that the spread in the abundances of our pairs shown in Figure \ref{fig:abundance_genuine} is due principally to measurement imprecision.

We therefore conclude that chemistry may be an effective way to separate wide binaries from random alignments, though more precise elemental abundances than those available in RAVE are required. However, any future study considering the chemistry of wide binaries should be cognizant of the non-negligible presence of {\it bona fide} wide binaries with different abundances (false negatives) and of random alignments with very similar chemistry (false positives).

These arguments largely follow our concern expressed in Paper I about the possibility of using RVs as a wide binary identification criterion: while requiring consistent RVs may be a powerful constraint used to identify wide binaries, interesting, {\it bona fide} wide binaries such as triple or higher order systems may cause a wide binary to have inconsistent RVs. Conversely, some 20\% of random alignments have consistent RVs, depending on measurement precision. Such a criterion may be extremely valuable but should be used with caution.

\subsection{What are the implications for pairs with pc-scale separations?}
\label{sec:pc_scale}

\begin{figure*}
	\includegraphics[width=\textwidth]{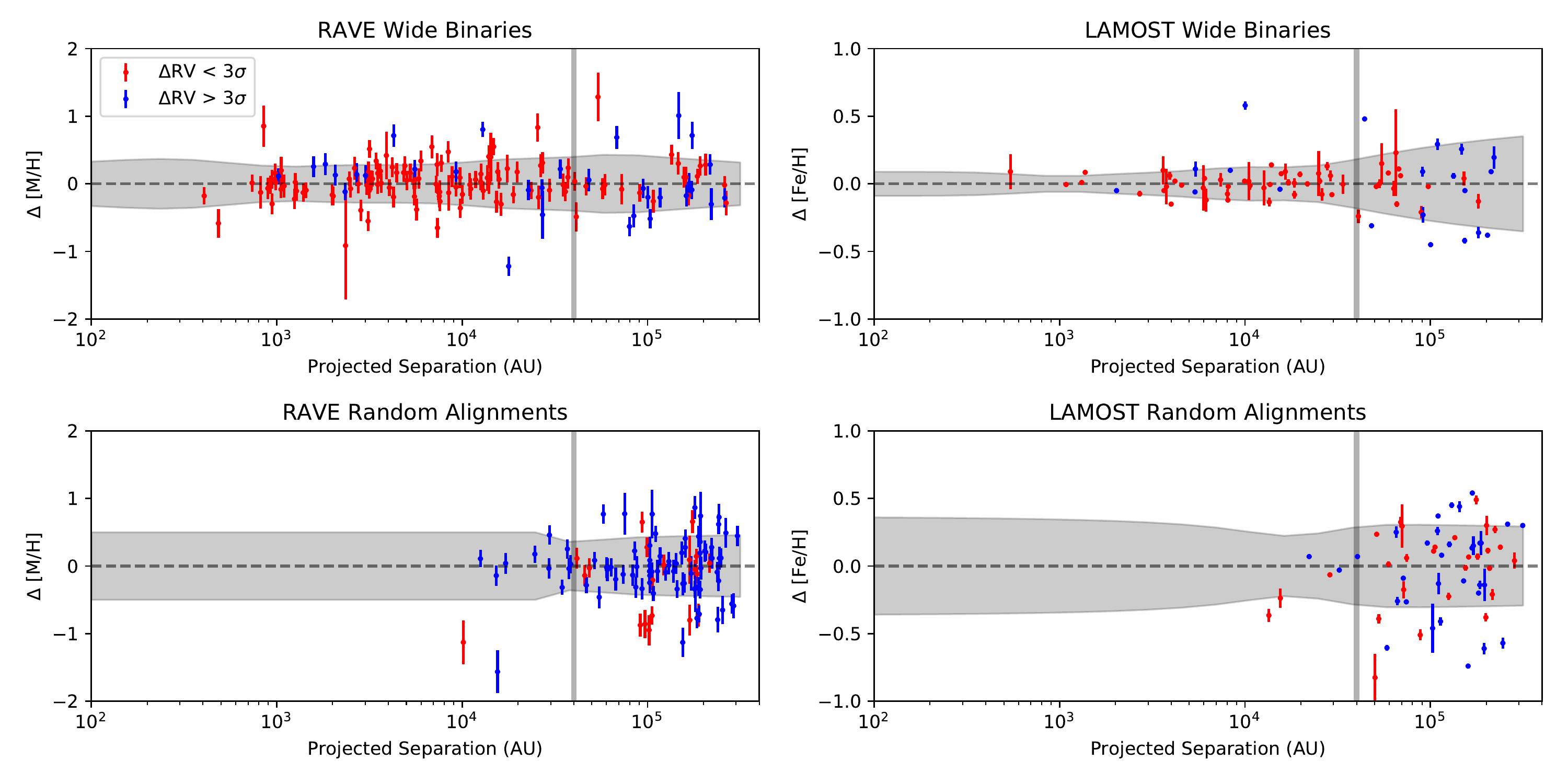}
    \caption{Differences in the metallicities of the components of genuine pairs in RAVE (top left) and LAMOST (top right) and random alignments in RAVE(bottom left) and LAMOST (bottom right). Systems in which the components have consistent (inconsistent) RVs at the 3$\sigma$ level are shown in red (blue). Grey backgrounds depict the running standard deviation of the pairs in each sample (it should not be confused with the run of typical metallicity measurement uncertainties). It is apparent from the LAMOST data, but not necessarily the RAVE data, that stellar pairs with s~$>$~4$\times$10$^4$AU, typically have inconsistent metallicities. The bottom panels show that random alignments become common at separations larger than a few 10$^4$ AU, and indeed these typically have inconsistent metallicities. Random alignments at smaller separations may exist, but are rare. Grey backgrounds show that the spread of the distributions of our catalog of wide binaries are similar to the spread of random alignments at separations of $\sim$10$^5$ AU. Note the different scale and range for the y-axis for the left and right panels. }
    \label{fig:proj_sep_delta_metallicity}
\end{figure*}

Previous searches for wide binaries and co-moving stars within TGAS have claimed to identify many stellar pairs with separations $\gtrsim$1 pc \citep{oelkers17, oh17}. In Paper I, we identified 4$\times$10$^4$ AU ($\approx$0.2 pc) as the separation limiting genuine binaries from a region where random alignments of unassociated stars were dominant, thus complicating the identification of genuine wide binaries based only on TGAS data\footnote{Wide binaries may exist at larger separations, but they become extremely difficult to robustly identify. For instance, in a recent RV follow-up preprint that was posted while the present manuscript was under review, \citet{price-whelan17} found that over half of the \citet{oh17} sample of stellar pairs at parsec-scale separations is comprised of random alignments.}. We showed that pairs at larger separations typically have inconsistent RVs, indicative of random alignments. If, instead, such pairs were truly associated (i.e., reflecting a common birth place), their metallicities should exhibit consistency as well. We perform this additional test next.

Figure~\ref{fig:proj_sep_delta_metallicity} compares the metallicity measurements of our sample of binaries as a function of projected separation for both RAVE (top left panel) and LAMOST stars (top right panel). Grey backgrounds depict the running standard deviation of the pairs in both samples. For pairs with large separations ($>$4$\times$10$^4$ AU; grey vertical line), metallicities measured by RAVE have only a slight variation in the scatter in metallicity differences. This is due to large uncertainties in the RAVE data; the top right panel, with more precise LAMOST metallicities, shows a clear increase in the scatter of metallicity differences at separations beyond 4$\times$10$^4$ AU (note the different scales on the y-axis).

The bottom panels in Figure~\ref{fig:proj_sep_delta_metallicity} show the corresponding metallicity comparison as a function of project separation for our set of random alignments. The distribution indicates that random alignments are possible, but not common at projected separations less than $\sim$10$^4$ AU (projected separation is already a powerful discriminant separating wide binaries from random alignments). There are relatively more random alignments at larger projected separations compared with our samples of wide binaries since we have used a somewhat looser criterion, selecting random alignments with a posterior probability above 50\% rather than 99\% as for our wide binaries. However, note that these pairs still have similar proper motions and parallaxes. Grey backgrounds indicate that the spread of the distributions of our samples of genuine wide binaries increases with projected separation and reaches a similar scale to the spread of random alignments at separations of $\sim$10$^5$ AU.

In Paper I, we argue based on our independent random alignment tests as well as our comparison using RVs, that the sample of pairs with separations larger than $\approx$4$\times 10^4$ AU (in our sample as well as others) are likely to be dominated by random alignments of unassociated stars. The metallicity differences in Figure \ref{fig:proj_sep_delta_metallicity} appear to support our claim: Not only do stellar pairs at separations of a few 0.1 pc and larger tend to have inconsistent RVs, but their metallicities tend to be inconsistent as well. Therefore, the bulk of these pairs (even those few with near-zero $\Delta$[Fe/H]) cannot be claimed to be the components of former wide binaries that were ionized, which were predicted by \citet{jiang10}. Instead, it is more likely that the stars in such widely separated pairs may be chance alignments of unassociated stars or members of larger, dynamically formed over-densities in the local Galactic neighbourhood such as those recently identified by \citet{antoja17} and \citet{liang17}. Although we will have to wait until future {\it Gaia} data releases with more precise astrometry, the current data indicate that pairs with such wide separations, even those with very similar astrometry, are overwhelmingly comprised of randomly aligned stars, as Figure \ref{fig:proj_sep_delta_metallicity} suggests.

\section{Wide binaries as tests of chemical tagging}
\label{sec:chemical_tagging}

\begin{figure}
	\includegraphics[width=\columnwidth]{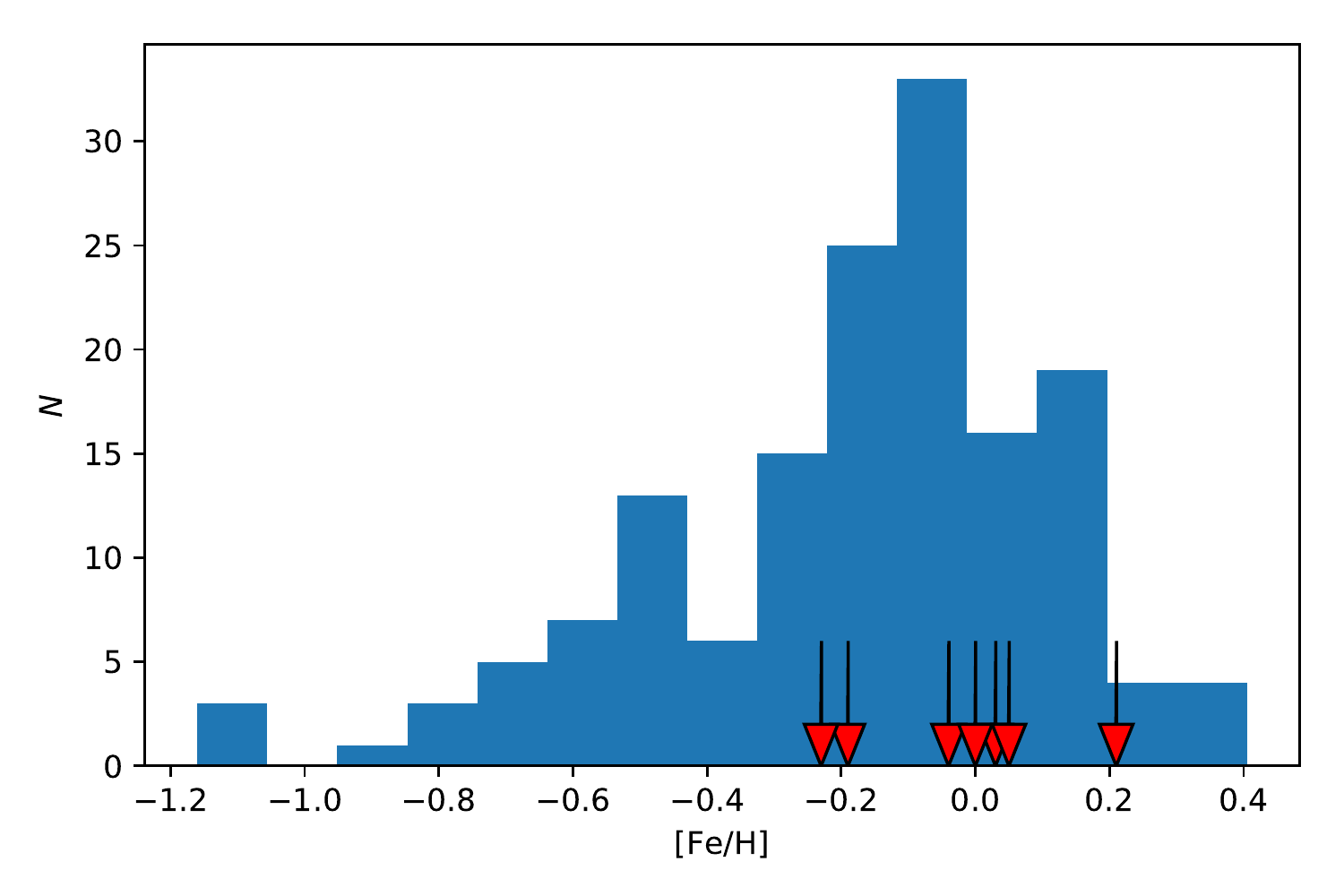}
    \caption{ The distribution of metallicities of the binaries in our sample as measured either by RAVE or LAMOST. Note that RAVE and LAMOST calibrate their metallicity measurements differently. We only include those pairs with RVs consistent at 3$\sigma$. Arrows show the metallicities of the seven open clusters containing the stars used to test chemical tagging by \citet{ness17}. The stars in our wide binaries from TGAS extend to both higher and, especially, lower metallicities than nearby open clusters. }
    \label{fig:metallicity_distribution}
\end{figure}

Tests of chemical tagging have typically been limited to stars within open clusters \citep{blanco-cuaresma15,bovy16,lambert16,ness17}. The utility of these stellar samples to test chemical tagging is constrained by the limited number of well characterized clusters, contamination by field stars, and source confusion within dense stellar systems. Furthermore, nearby, well-studied open clusters are typically young and therefore have near-solar metallicities. In the most expansive test to date of chemical tagging, \citet{ness17} use stars in seven open clusters with [Fe/H] ranging from -0.23 to $+$0.24. The red arrows in Figure \ref{fig:metallicity_distribution} indicate the metallicities of these clusters.

As an alternative test, \citet{hogg16} use a clustering algorithm to find that stars with similar abundances aggregate in phase space. The results from this study are extremely encouraging, although this test, too, has its limitations, as it can be difficult to interpret. For instance, \citet{hogg16} identify a structure in abundance space that may be associated with the globular cluster M53, but confirmation requires a substantial observational effort.

Here we put forward the idea that catalogs of reliably classified wide binaries form a unique sample to calibrate chemical tagging as a method to identify substructure within the Galaxy. The metallicity and elemental abundance consistency between wide binary components presented in this work bolsters the status of wide binaries as ``mini-open clusters,'' and therefore potential samples with which to test chemical tagging.

One major benefit of wide binaries is that they span a range in [Fe/H] from -1.0 to $+$0.5. Figure \ref{fig:metallicity_distribution} shows that range is far wider, particularly at low metallicities, than what is spanned by the open clusters used in \citet{ness17}.

The metallicity distribution of the current sample of {\it Gaia} wide binaries depicted in Figure \ref{fig:metallicity_distribution} is an artifact of the combination of the non-trivial selection functions for TGAS, RAVE, and LAMOST stars. With an expansion of orders of magnitude in size, upcoming {\it Gaia} data releases will facilitate the identification of even larger wide binary samples with a metallicity coverage improved over the current sample. Of particular interest would be to test the regime of metal poor stellar populations, with metallicities typical of those of the Galactic stellar halo. These data would provide the calibration necessary to determine if large structures in abundance space, such as that identified by \citet{hogg16} that may be associated with M53, are genuine or merely associated in abundance space by chance.

As an additional example, using more precise abundances than what is available from RAVE, wide binaries can test the presence of unassociated stars that may appear to have identical chemistry. Naming such pairs doppelgangers, \citet{ness17} demonstrate that $\sim$1\% of stars with similar metallicities have elemental abundances within 0.03 dex.

Wide binaries can build off the work of \citet{ness17} by testing the presence of a potentially more pernicious source of contamination for chemical tagging studies: pairs of unassociated stars with very similar abundances across multiple elements {\it and} similar kinematics. In Paper I, our tests with random alignments indicate that there exists a non-negligible number of unassociated stars with similar proper motions and parallaxes (Of course, if such stars occupy nearly identical positions in all six dimensions of phase space, they must necessarily be associated). These arise from uncertainties in the astrometry (particularly in the parallax) which can lead to confusion between genuinely associated pairs and randomly aligned stars with similar positions in phase space. 

Here, we find the first indications that some fraction of these unassociated stellar pairs with similar kinematics may also have similar chemistry. As mentioned in Section \ref{sec:abundances_for_matching}, even with precise LAMOST metallicities, a third of our sample of random alignments have metallicities consistent at 3$\sigma$. A more stringent constraint requiring 1$\sigma$ consistency in the metallicity reduces the contamination rate to $\sim$10\%. Requiring consistency of multiple elemental abundances will likely reduce this rate even further (as we discuss in Section \ref{sec:abundances_for_matching}, RAVE abundance measurements lack the precision required to be powerful classifiers, except for possibly at significantly sub-solar or super-solar metallicities). In addition to chemistry, some fraction of these random alignments also have radial velocities consistent at the 3$\sigma$ level, as can be seen by the presence of red points in the bottom panels of Figures \ref{fig:metallicity} and \ref{fig:proj_sep_delta_metallicity}. Depending on their rate, such chemical and kinematic doppelgangers could prove problematic for chemical tagging studies.

Elemental abundances more precise than those available in RAVE are required to go beyond the present discussion. A detailed test requires high-resolution and signal-to-noise ratio spectra of the wide binaries in our sample which would provide a statistical measure of the expected abundance differences between related stars. At the same time, equivalent spectra of the stars in our corresponding catalog of random alignments would constrain the rate of chemical and kinematic doppelgangers. We leave this for future work.

\section{Conclusions}
\label{sec:conclusions}

From our catalog of 7108 TGAS wide binaries, we select the subset of pairs in which the posterior probability is above 99\% when using a power-law prior on the distribution of orbital separations. To reduce contamination from randomly aligned stellar pairs, we select only those pairs with projected separations less than 4$\times$10$^4$ AU, which winnows our sample down to 4491 binaries. The contamination rate of these pairs should be $\approx$5\%. We cross-match these pairs with the RAVE DR5 catalog \citep{kunder17} and the  LAMOST DR3 catalog and obtain 115 wide binaries in which both components are detected by RAVE and 62 with LAMOST data.

\begin{enumerate}
\item We compare the metallicities between the two components of wide binaries by cross-matching our sample of binaries with RAVE and LAMOST. Comparison of the metallicities between the two components of our binaries show a significant positive correlation, a property not seen in the corresponding sample of randomly aligned stars. The differences in these metallicities are consistent with the precision limit of the available data.
\item We compare the abundances of Mg, Al, Si, Ti, and Fe (directly from the Fe lines) of the two stars in each binary in our sample. These elements all show significant positive correlations, with elemental abundances of the components of wide binaries consistent with observational precision.
\item The metallicity and abundance comparisons indicate that individual binaries were formed from the same parent material, potentially stellar clusters. The vast majority of wide binaries cannot be formed dynamically either in the field or in local stellar overdensities caused by Galactic structure. Wide binaries are co-chemical and therefore it is reasonable to assume they are co-eval. This justifies their use to calibrate age-activity-rotation relations, initial-final mass relation for white dwarfs, and M-dwarf metallicity indicators.
\item RAVE derived abundances and metallicities are not precise enough to separate {\it bona fide} wide binaries from random alignments. In conjunction with kinematics, the more precise LAMOST DR3 metallicities can likely aid in the identification of binaries. Future, precise spectroscopic surveys may allow for chemistry to be included as a matching criterion; however this criterion should be used with caution.
\item The co-chemical nature of wide binaries indicates they may be a data set complementary to open cluster stars by which to test the viability of chemical tagging, particularly because wide binaries span a wider range in metallicity compared with open clusters.
\item Common proper motion, common parallax stellar pairs in TGAS at projected separations $\sim$pc typically have inconsistent metallicities indicating they are predominantly comprised of randomly aligned, unassociated stars, in agreement with our results from Paper I.
\end{enumerate}

\section*{Acknowledgements}

We thank the anonymous referee for helpful comments and suggestions. We also thank Phillip Cargile, Melissa Ness, Keith Hawkins, and Simon Schuler for comments on an earlier draft of the manuscript. J.J.A. acknowledges funding from the European Research Council under the European Union's Seventh Framework Programme (FP/2007-2013)/ERC Grant Agreement n. 617001. J.C. acknowledges support from the Chilean Ministry for the Economy, Development, and Tourism's Programa Iniciativa Cient\'{i}fica Milenio, through grant IC120009 awarded to the Millenium Institute of Astrophysics (MAS), from PFB-06 Centro de Astronomia y Tecnologias Afines.

This work was performed in part at Aspen Center for Physics, which is supported by National Science Foundation grant PHY-1607611. This work was partially supported by a grant from the Simons Foundation.

Guoshoujing Telescope (the Large Sky Area Multi-Object Fiber Spectroscopic Telescope LAMOST) is a National Major Scientific Project built by the Chinese Academy of Sciences. Funding for the project has been provided by the National Development and Reform Commission. LAMOST is operated and managed by the National Astronomical Observatories, Chinese Academy of Sciences.

Funding for RAVE has been provided by: the Australian Astronomical Observatory; the Leibniz-Institut fuer Astrophysik Potsdam (AIP); the Australian National University; the Australian Research Council; the French National Research Agency; the German Research Foundation (SPP 1177 and SFB 881); the European Research Council (ERC-StG 240271 Galactica); the Istituto Nazionale di Astrofisica at Padova; The Johns Hopkins University; the National Science Foundation of the USA (AST-0908326); the W. M. Keck foundation; the Macquarie University; the Netherlands Research School for Astronomy; the Natural Sciences and Engineering Research Council of Canada; the Slovenian Research Agency; the Swiss National Science Foundation; the Science \& Technology Facilities Council of the UK; Opticon; Strasbourg Observatory; and the Universities of Groningen, Heidelberg and Sydney. The RAVE web site is at https://www.rave-survey.org.

%%%%%%%%%%%%%%%%%%%%%%%%%%%%%%%%%%%%%%%%%%%%%%%%%%


\bibliographystyle{mnras}
\bibliography{references} 

%%%%%%%%%%%%%%%%%%%%%%%%%%%%%%%%%%%%%%%%%%%%%%%%%%

% Don't change these lines
\bsp	% typesetting comment
\label{lastpage}
\end{document}